\documentclass[conference,10pt]{IEEEtran}
\IEEEoverridecommandlockouts
\usepackage[utf8]{inputenc}
\usepackage{graphicx} 
\usepackage{bm}
\usepackage{amsmath}
\usepackage{amsfonts}
\usepackage{algorithm2e}
\usepackage{algpseudocode}
\usepackage{multirow}
\usepackage{amssymb}
\usepackage{pifont}
\usepackage{subfigure}
\usepackage{soul}
\usepackage{tabularx}
\newcolumntype{Y}{>{\centering\arraybackslash}X}
\newcolumntype{K}[1]{>{\centering\arraybackslash}m{#1}}
\newcolumntype{L}[1]{>{\centering\arraybackslash}p{#1}}
\usepackage{multirow}
\usepackage{xcolor}
\usepackage{marginnote}
\usepackage{subfigure}
\usepackage{cite}
\usepackage{bm}
\usepackage{mathtools}
\newcommand{\norm}[1]{\left\lVert#1\right\rVert}
\author{
    \IEEEauthorblockN{Zihang Song, Matteo Zecchin, Bipin Rajendran and Osvaldo Simeone}
    \IEEEauthorblockA{Centre for Intelligent Information Processing Systems (CIIPS)\\Department of Engineering \\
King’s College London\\
London, UK}
\thanks{
	The authors are with the Centre for Intelligent Information Processing Systems (CIIPS), Department of Engineering, King’s College London, London WC2R 2LS, U.K. (e-mail: \{zihang.song, matteo.1.zecchin, bipin.rajendran, osvaldo.simeone\}@kcl.ac.uk).
	
	This work is supported in part by the European Union's Horizon Europe project CENTRIC (101096379), the EPSRC project (EP/X011852/1), and the Open Fellowships of the EPSRC (EP/W024101/1 and EP/X011356/1).}
}

\DeclareMathAlphabet\mathbfcal{OMS}{cmsy}{b}{n}
\usepackage{titlesec}

\begin{document}
\title{In-Context Learned Equalization in Cell-Free Massive MIMO via State-Space Models}
\maketitle

\begin{abstract}
 Sequence models have demonstrated the ability to perform tasks like channel equalization and symbol detection by automatically adapting to current channel conditions. This is done without requiring any explicit optimization and by leveraging not only short pilot sequences but also contextual information such as long-term channel statistics. The operating principle underlying automatic adaptation is in-context learning (ICL), an emerging property of sequence models. Prior art adopted transformer-based sequence models, which, however,  have a computational complexity scaling quadratically with the context length due to batch processing. Recently, state-space models (SSMs) have emerged as a more efficient alternative, affording a linear inference complexity in the context size.  This work explores the potential of SSMs for ICL-based equalization in cell-free massive MIMO systems.  Results show that selective SSMs achieve comparable performance to transformer-based models while requiring approximately eight times fewer parameters and five times fewer floating-point operations.

\end{abstract}

\begin{IEEEkeywords}
In-Context Learning, State-Space Models, Cell-Free Massive MIMO, Channel Equalization
\end{IEEEkeywords}

\section{Introduction}

\noindent \textbf{Context and motivation:}  Future-generation wireless networks, such as those based on the Open Radio Access Network (O-RAN) specification, will be characterized by native AI interfaces \cite{bonati2021intelligence}, enabling some of the network functionalities to be implemented through AI modules integrated into  RAN Intelligent Controllers (RICs) \cite{balasubramanian2021ric}. In the face of dynamic channel and network conditions, the effectiveness of AI-based algorithms hinges on their ability to adapt based on limited contextual information, such as pilot sequences, topological data, and channel statistics, to deliver consistent and satisfactory services \cite{hou2024automaticaimodelselection}.

Recently, \emph{in-context learning} (ICL) \cite{garg2022can} has emerged as a promising paradigm to realize AI algorithms that can adapt autonomously based on contextual information   \cite{zecchin2023context,rajagopalan2023transformers,zecchin2024cell,song2024neuromorphic,abbas2024leveraging,song2024transformer}. The primary benefit of ICL is that adaptation occurs without the need for explicit optimization and model fine-tuning.  A sequence model implementing ICL takes as input a sequence of transmitted and received pilot signals, along with any other contextual information, to map a new input to a target variable.  ICL was shown to offer major advantages over existing methods like meta-learning \cite{chen2023learning,simeone2022machine}, which instead achieve adaptation by updating an AI model’s parameters \cite{chen2023learning,nikoloska2022modular}.

\begin{figure}[t]
    \centering
    \includegraphics[width=0.9\linewidth]{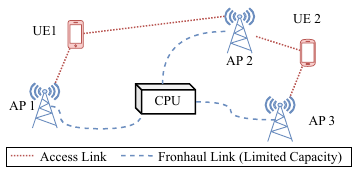}
    \caption{A cell-free MIMO setup with user equipment (UEs) communicating to distributed access points (APs) that are coordinated by a central processing unit (CPU). The APs are connected to the CPU via fronthaul links with limited capacity, which are represented as dashed blue lines. }
    \label{fig:cellfreemimo}
\end{figure}

Cell-free multi-input and multiple-output (MIMO) is a networking architecture compatible with the functionality split adopted by O-RAN systems. As shown in Fig. \ref{fig:cellfreemimo}, this architecture involves a central processing unit (CPU) that processes signals received from access points (APs) distributed throughout a designated deployment area \cite{quek2017cloud,nayebi2015cell,bjornson2019making}. Communication between the APs and the CPU occurs over limited-capacity fronthaul links  \cite{ibrahim2021uplink,masoumi2019performance}. For this problem, transformer-based sequence \cite{Vaswani2017AttentionIA} models trained via ICL have demonstrated the ability to outperform existing model-based solutions by mitigating the effects of limited fronthaul capacity and pilot contamination based on a few pilots  \cite{zecchin2024cell}.

While in principle ICL-equalizers can be realized via any sequence model implementing ICL, existing work has been limited so far to the class of transformer-based sequence models  \cite{zecchin2023context,rajagopalan2023transformers,zecchin2024cell,song2024neuromorphic}. Despite their success, these models are known to have high computational requirements, which scale quadratically with the context length due to batch processing.  More recently, \emph{state-space models} (SSMs) have emerged as a promising class of efficient ICL-capable sequence models  \cite{gu2021efficiently,gu2023mamba}. Unlike transformers, SSMs can process sequences sequentially, thereby ensuring a linear scaling of the computational complexity relative to the context size. While works \cite{sushma2024state,park2024can} have also demonstrated the ability of SSMs to perform ICL, the potential of SSMs for ICL-based channel equalization has yet to be explored.

\noindent \textbf{Contributions:} This work presents an SSM-based ICL (SSM-ICL) model that implements a lightweight channel equalizer for multi-user cell-free MIMO systems. The proposed model efficiently leverages transmitted and received pilot sequences, along with channel statistics, as contextual information to carry out equalization. Specifically, the key contributions are as follows:
\begin{itemize} 
\item We introduce  SSM-ICL, an equalizer for multi-user cell-free MIMO systems with constrained fronthaul capacity that can automatically adapt to channel conditions. SSM-ICL employs efficient sequential data processing, providing a computationally efficient alternative to transformer-based attention mechanisms.
\item Numerical results show that SSM-ICL equalizers not only outperform model-based channel equalizers, but they also achieve comparable performance to transformer-based ICL equalizers, while using up to $8.5\times$ fewer parameters and reducing floating-point operations by a factor of $5.5$.
\end{itemize}

\section{Background: State-Space Models}
\label{sec:ssm_overview}
In this section, we focus on \textit{scalar} input-output SSMs, which map a continuous-time input process $u(t)\in \mathbb{R}$ into an output process $o(t)\in \mathbb{R}$. This is done by modeling their relation via the continuous-time dynamics of a hidden state $\bm{h}(t) \in \mathbb{R}^{N\times 1}$. The hidden state evolution is governed by the first-order differential equation
\begin{equation}
\label{eq:state_cont}
    \begin{aligned}
    \frac{\mathrm{d}\bm{h}}{\mathrm{d}t} &= \bm{A} \bm{h}(t) + \bm{b} u(t),
    \end{aligned}
\end{equation}
where $\bm{A} \in \mathbb{R}^{N \times N}$ is the state transition matrix and $\bm{b} \in \mathbb{R}^{N\times 1}$ is a vector.
The output process $o(t)$ is obtained by projecting the hidden state process $\bm{h}(t)$  along a vector $\bm{c} \in \mathbb{R}^{1 \times N}$ as
\begin{equation}
\label{eq:output_cont}
    \begin{aligned}
    o (t)&= \bm{c}\bm{h}(t).
    \end{aligned}
\end{equation}

In practice, continuous-time dynamics are discretized by using a step size $\Delta$. In the following, we use subscript $t$ to indicate discrete time steps. The corresponding discrete-time equations are given by \cite{gu2021efficiently}
\begin{subequations}\label{eq:ssm}
\begin{align}
    \bm{h}_t &= \overline{\bm{A}}\bm{h}_{t-1}+\overline{\bm{b}}u_t, \label{eq_1}\\
    o_t &= \bm{c}\bm{h}_t, \label{eq_2}
\end{align}
\end{subequations}
where the model parameters are defined as
\begin{subequations}\label{eq:overlineAB}
    \begin{align}
        \overline{\bm{A}}&=(\bm{I}-\Delta/2\cdot \bm{A})^{-1}(\bm{I}+\Delta/2\cdot \bm{A}),\label{eq:overlineA}\\
        \overline{\bm{b}}&= (\bm{I}-\Delta/2\cdot \bm{A})^{-1}\Delta \bm{b}.\label{eq:overlineB}
    \end{align}
\end{subequations}
By replacing the continuous-time update equations \eqref{eq:state_cont}-\eqref{eq:output_cont} with its discrete-time counterparts \eqref{eq:ssm}, SSMs can be efficiently implemented either in the form of a convolution or as a recurrent model. The first approach supports parallel training of SSMs as for transformers. Furthermore, a recurrent implementation ensures an inference complexity that scales linearly with the context length, in contrast to the quadratic complexity of transformers.

The state transition matrix $\bm{A}$ is typically fixed, while both the vectors $\bm{b}$ and $\bm{c}$ are trainable parameters. Specifically, the matrix 
$\bm{A}$ can be designed to capture long-range dependencies. Notably, the structured SSM (S4) model chooses the lower-triangular matrix \cite{gu2021efficiently}
\begin{equation}
    A_{ij} = -\left\{
\begin{array}{ll}
\sqrt{(2i+1) (2j+1)} & \text{if } i > j, \\
i+1 & \text{if } i = j, \\
0 & \text{if } i < j,
\end{array}
\right.
\end{equation}
so that \( \bm{A} \) admits a normal-plus-low-rank representation. This structure is preserved in matrix \( \overline{\bm{A}} \), enabling a computationally efficient implementation for both parallel training and recurrent inference.

The capacity of the S4 model is constrained by the use of a time-invariant dynamic model \((\bm{A}, \bm{b}, \bm{c})\) and step size $\Delta$, which effectively treats each input token in the same way. To enable context-dependent reasoning, a selective SSM, known as \textit{Mamba}, was proposed in \cite{gu2023mamba}. Mamba adopts input-dependent vectors \( \bm{b}_t \) and \( \bm{c}_t \), as well as an input-dependent time step \( \Delta_t \), resulting in input-dependent parameters ($\overline{\bm{A}}_t, \overline{\bm{b}}_t$) in the discretized process \eqref{eq:overlineAB}. We refer to \cite{gu2023mamba} and Section \ref{sec:selective_ssm} for further details. 

\section{System Model}
\label{sec:system_model}
We consider the uplink of a cell-free network comprising of $K$ single-antenna UEs and $M$ APs, each equipped with $N_{\text{ant}}$ antennas.  The APs are connected to a CPU via fronthaul links with limited capacity. The CPU coordinates their operations and processes the received signals. In 5G or O-RAN, the APs correspond to radio units, while CPUs are implemented as the distributed units \cite{demir2024cell}.

The wireless channel between each UE $k$ and each AP $m$ 
is denoted by the vector $\bm{h}_{m,k}\in\mathbb{C}^{N_{\text{ant}}\times1}$, which follows a Rayleigh block-fading channel model. The channel vector  $\bm{h}_{m,k}$ is assumed constant within each coherence block of $T$ channel uses, and it follows a correlated Rayleigh distribution
\begin{align}
    \bm{h}_{m,k}\sim \mathcal{CN}(\bm{0},\bm{R}_{m,k}),
    \label{eq:rayleigh_ch}
\end{align}
where $\bm{R}_{m,k}$ is the $N_{\text{ant}}\times N_{\text{ant}}$ spatial correlation matrix. The large-scale fading coefficient 
\begin{align}
    r_{m,k}=\textrm{tr}(\bm{R}_{m,k})/N_{\text{ant}}
\end{align}
 measures the average channel gain from user $k$ to AP $m$. The large-scale fading coefficients for the channels from UE $k$ to all APs are collected in vector  $\bm{r}_{k}=[r_{1,k},\dots,r_{m,k}]$.

The unknown channel $\bm{h}_{m,k}$ between each UE $k$ and AP $m$ is estimated via pilot transmission. During the first  $T_{\text{P}}< T$  channel uses of each coherence block, the UEs transmit known pilot sequences, which are used for channel estimation. The remaining  $T - T_{\text{P}}$  channel uses are reserved for data transmission.  During the pilot transmission phase, each UE $k$ is assigned a pilot sequence 
\begin{equation}
    \bm{x}^{\text{P}}_{k} = \left[x^{\text{P}}_{k,1},\ldots,x^{\text{P}}_{k,T_{\text{P}}}\right]
\end{equation}
encompassing $T_{\text{P}}$ symbols, with normalized energy  $\norm{x^{\text{P}}_{k}}^2 = T_{\text{P}}$. The pilot sequence $\bm{x}^{\text{P}}_{k}$ is chosen from a set $\{\bm{x}^{\text{P}}_{1}, \dots, \bm{x}^{\text{P}}_{T_{\text{P}}}\}$ of $T_{\text{P}}$ orthogonal pilot sequences. As in standard cell-free massive MIMO systems \cite{nayebi2015cell},  pilot reuse is allowed, with different UEs possibly being assigned the same pilot sequence. The set of UEs sharing the same pilot as UE $k$, excluding $k$ itself, is denoted as 
\begin{equation}
    \mathcal{U}_k = \{ j \in \{1, \dots, K\} \mid j \neq k, \bm{x}^{\text{P}}_{j} = \bm{x}^{\text{P}}_{k}  \}.
\end{equation}

The received signal at the $m$-th AP during the pilot transmission phase is given by 
\begin{align}
    \bm{Y}^{\text{P}}_m=\sum^K_{k=1}\bm{H}_{m,k}{\bm{x}^{\text{P}}_{k}}+\bm{N}_m,
    \label{eq:received_pilots}
\end{align}
where $\bm{N}_m\in \mathbb{C}^{N_{\text{ant}}\times T_{\text{P}}}$ is the receiver noise matrix comprising independent complex Gaussian entries $\left[\bm{N }^{}_m\right]_{i,j}\sim \mathcal{CN}(0,\sigma^2)$.

In the data transmission phase, each UE $k$ transmits a data symbol $x_k\in\mathcal{X}_k$ chosen uniformly at random from a constellation $\mathcal{X}_k$ with unit average transmit power; i.e., $\mathbb{E}[\norm{x_k}^2]=1$. The concatenation of the transmitted data symbols is denoted as
$\bm{x}=\left[x_1,\dots,x_K\right]^{\mathsf{T}}\in \mathbb{C}^{K}$.  
The received signal at AP $m$ during each channel use of the data transmission phase is given by
\begin{align}
     \bm{y}_m=\sum^K_{k=1}\bm{H}_{m,k}x_k+\bm{n}_m
    \label{eq:received_data}
\end{align}
where $\bm{n}_m\sim  \mathcal{CN}(\bm{0},\sigma^2\bm{I})$ is the receiver noise vector.

Following the common 7.2x fronthaul split \cite{3gpp_split}, channel equalization is carried out at the CPU, with each AP $m$ acting as a relay that forwards the received pilot signals  $\bm{Y}^{\text{P}}_m$  and the data signals $\bm{y}_m$  to the CPU for processing.

The fronthaul links between the APs and the CPU have a capacity of $b$ bits per symbol of the wireless interface. To meet the fronthaul capacity constraints, each AP $m$ applies a scalar $b$-bit quantizer $\mathcal{Q}_b(\cdot)$ separately to the real and imaginary components of the baseband signals $\bm{Y}^{\text{P}}_m$ and $\bm{y}_m$. The quantized received pilot and data signals are denoted by $\tilde{\bm{Y}}^{\text{P}}_m$ and $\tilde{\bm{y}}_m$ respectively.  
The received quantized pilots at the CPU from all $M$ APs are concatenated row-wise as $\tilde{\mathbfcal{Y}}^{\text{P}}=\big[[\tilde{\bm{Y}}^{\text{P}}_1]^{\mathsf{T}},\ldots,[\tilde{\bm{Y}}^{\text{P}}_M]^{\mathsf{T}}\big]^{\mathsf{T}}\in \mathbb{C}^{N_{\text{ant}} M\times T_{\text{P}}}$. 
In a similar way, the concatenation of all quantized received data is denoted as $\tilde{\bm{y}}=[{\tilde{\bm{y}}_1}^{\mathsf{T}},\ldots,{\tilde{\bm{y}}_M}^{\mathsf{T}}]^{\mathsf{T}}\in \mathbb{C}^{N_{\text{ant}} M\times 1}$.

Overall,  a channel equalization task $\tau$ is specified by the tuple 
\begin{align}
    \label{eq:task_info}
    \tau=\left(\sigma^2, K, \{\mathcal{X}_k\}^K_{k=1},\{\bm{R}_{1,k}\}^K_{k=1},\dots, \{\bm{R}_{m,k}\}^K_{k=1}\right),
\end{align} 
which determines the noise variance $\sigma^2$, the number $K$ of active UEs, the constellations $\{\mathcal{X}_k\}^K_{k=1}$,  and the spatial correlation matrices $\{\bm{R}_{1,k}\}^K_{k=1},\dots, \{\bm{R}_{m,k}\}^K_{k=1}$ of the channels.

\section{SSM-based ICL Equalizer }

\begin{figure}
    \centering
    \includegraphics[width=0.9\linewidth]{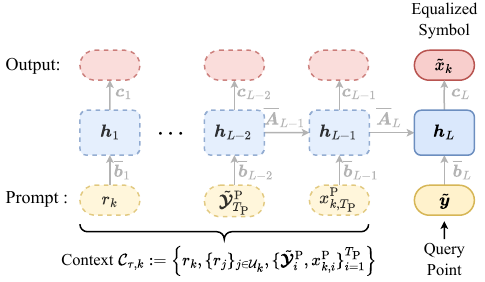}
    \caption{Illustration of a 1-layer SSM-ICL equalizer, where $\overline{\bm{A}}_l$ and $\overline{\bm{b}}_l$ represents the collection of parameters $\overline{\bm{A}}_{l,d}$ and $\overline{\bm{b}}_{l,d}$ for each parallel index by $d=1,\ldots,D$.}
    \label{fig:ssm-icl}
\end{figure}

In this section, we introduce the SSM-based ICL (SSM-ICL) equalizer, which leverages the Mamba SSM model \cite{gu2023mamba} to estimate the transmitted symbols \( \bm{x} = \left[x_1, \dots, x_K\right]^{\mathsf{T}} \in \mathbb{C}^K \) based on the transmitted pilots \( \{\bm{x}_k^{\text{P}}\}_{k=1}^K \), the quantized received pilots \( \tilde{\mathbfcal{Y}}^{\text{P}} \), and the received quantized signal \( \tilde{\bm{y}} \). The proposed architecture is illustrated in Fig. \ref{fig:ssm-icl}.

The spatial correlation matrices \( \bm{R}_{m,k} \) and the noise variance \( \sigma^2 \) are unknown at the CPU. Thus, the CPU has only limited access to the task information \( \tau \) in \eqref{eq:task_info}. The available information, denoted as \( f(\tau) = \{\{\bm{r}_k\}_{k=1}^K, \{\mathcal{X}_k\}_{k=1}^K\} \), includes the large-scale fading coefficients \( \{\bm{r}_k\}_{k=1}^K \), and the UE data constellations \( \{\mathcal{X}_k\}_{k=1}^K \). The SSM-ICL equalizer employs parallel decoding, whereby decoding for each UE $k$ is carried out separately. In the following, we demonstrate the equalization process for the \( k \)-th UE as an example.

\subsection{Prompt Design}
As in \eqref{eq:rayleigh_ch}, to equalize the data symbol $x_k$ for UE $k$, the task information $f(\tau)$ is first mapped into a user-dependent context $\mathcal{C}_{\tau,k}$ along with pilots and the received signal $\tilde{\bm{y}}$. Accordingly, the context is defined as
\begin{equation}\label{eq:context}
    \mathcal{C}_{\tau,k} \coloneq \left\{\bm{r}_k,\{\bm{r}_j\}_{j \in \mathcal{U}_k},\{\mathbfcal{\tilde{Y}}^{\text{P}}_i,x^{\text{P}}_{k,i}\}^{T_{\text{P}}}_{i=1}\right\},
\end{equation}
where $\mathbfcal{\tilde{Y}}^{\text{P}}_i$ denote the $i$-th column of matrix $\tilde{\mathbfcal{Y}}^{\text{P}}$. 
The context $\mathcal{C}_{\tau,k}$ in \eqref{eq:context} contains the large-scale fading information $\bm{r}_k$ for the given user $k$, as well as the large-scale coefficients $\{\bm{r}_j\}_{j \in \mathcal{U}_k}$ for other UEs sharing the same pilot sequence as user  $k$. This provides the SSM with statistical information about the channels of the UEs affected by pilot contamination, potentially helping to mitigate the impact of interference during pilot transmission. The context \eqref{eq:context} also contains $T_{\text{P}}$ in-context examples $(\tilde{\mathbfcal{Y}}^{\text{P}}_i,x^{\text{P}}_{k,i})$.

The received quantized signal $\tilde{\bm{y}}$ corresponding to the transmitted signal $x_k$ to be estimated is appended to the context $\mathcal{C}_{\tau,k}$ to obtain the final prompt
\begin{equation}
    \mathcal{P}_{\tau,k}\coloneq\{\mathcal{C}_{\tau,k},\tilde{\bm{y}}\}.
\end{equation}

\subsection{Tokenization}
The prompt $\mathcal{P}_{\tau,k}$ contains scalars $\{x^{\text{P}}_{k,i}\}^{T_{\text{P}}}_{i=1}$ and vectors $\bm{r}_k,\{\bm{r}_j\}_{j \in \mathcal{U}_k}, \{\mathbfcal{\tilde{Y}}^{\text{P}}_i\}^{T_{\text{P}}}_{i=1}$ and $\tilde{\bm{y}}$ of different lengths, which are converted to a sequence of input tokens via a tokenization process. Specifically, complex-valued scalars and vectors in the prompt are converted to real-valued tokens by separating the real and imaginary components. 

Additionally, in order to enable modulation-aware estimation, the received quantized signal $\tilde{\bm{y}}$ is augmented with information about the known constellation $\mathcal{X}_k$ adopted by the UE $k$. Given $N_{\text{mod}}$ possible constellations, each constellation $\mathcal{X}_k$ is is mapped to an index $\text{idx}(\mathcal{X}_k) \in \{0, \ldots, N_{\text{mod}}-1\}$, which is concatenated with $\tilde{\bm{y}}$ to create the token vector $\tilde{\bm{y}}^+ = [\tilde{\bm{y}}, \text{idx}(\mathcal{X}_k)]$. 

Finally, zero padding is applied to ensure that all token vectors match the length of the longest vector in the prompt which is denoted as $D$. 
The tokenization process results in an input token sequence $\bm{u}_1,\ldots,\bm{u}_L\in\mathbb{R}^{D}$, where $\bm{u}_1,\ldots, \bm{u}_{L-1}$ correspond to tokenized context vectors and $\bm{u}_L$ correspond to the tokenized query point. 

\subsection{Selective SSM}
\label{sec:selective_ssm}
The token sequence \( \bm{u}_1, \ldots, \bm{u}_L \in\mathbb{R}^D\) is processed using the Mamba SSM architecture \cite{gu2023mamba}, which recursively applies $N_{\text{layer}}$ selective SSM layers as detailed in Algorithm \ref{al:s6}. Specifically, each $d$-th scalar component in the input token vector $\bm{u}_l$ is used to update an individual discrete-time version of the state-space process in \eqref{eq:ssm}. This process is performed in parallel for  all entries $d\in\{1,\ldots,D\}$, and the output scalars are re-concatenated to generate the $l$-th output token as $\bm{o}_l=[o_{l,1},\ldots,o_{l,D}]^{\mathsf{T}}$.

\RestyleAlgo{ruled}
\SetAlgoLined
\begin{algorithm}\label{al:s6}
\LinesNumbered
\caption{Selective SSM Layer (Mamba)}
\KwIn{$\bm{u}_1,\ldots,\bm{u}_L\in\mathbb{R}^{D}$}
\KwOut{$\bm{o}_1,\ldots,\bm{o}_L\in\mathbb{R}^{D}$}
\For{$l = 1$ \KwTo $L$}{
    $\bm{b}_l, \bm{c}_l\in\mathbb{R}^{N} \gets \text{Linear}_N(\bm{u}_l)$\;
    $\bm{\Delta}_l\in\mathbb{R}^{D}\hspace{-0.2em} \gets\hspace{-0.15em}  \text{Softplus}(\epsilon\hspace{-0.075em} + \hspace{-0.075em}(\text{Linear}_1(\bm{u}_l),\dots,\text{Linear}_1(\bm{u}_l)))$\;
    \ForPar{$d = 1$ \KwTo $D$}{
    $\overline{\bm{A}}_{l,d}=(\bm{I}-\Delta_{l,d}/2\cdot \bm{A})^{-1}(\bm{I}+\Delta_{l,d}/2\cdot \bm{A})$
    \quad\hfill \Comment{\eqref{eq:overlineA}}\;
    $\overline{\bm{b}}_{l,d}=(\bm{I}-\Delta_{l,d}/2\cdot \bm{A})^{-1}\Delta_{l,d} \bm{b}_{l}$
    \Comment{\eqref{eq:overlineB}}\;
    $\bm{h}_{l,d} \gets \left\{
        \begin{array}{ll}
        \overline{\bm{A}}_{l,d}\bm{h}_{l-1,d} + \overline{\bm{b}}_{l,d}u_{l,d} & \text{if } l > 1 \\
        \overline{\bm{b}}_{l,d}u_{l,d} & \text{if } l = 1
        \end{array}\right. $\;
    $o_{l,d} \gets {\bm{c}_l}\bm{h}_{l,d}$\;
    }
$\bm{o}_l=[o_{l,1},\ldots,o_{l,D}]^{\mathsf{T}}$\;
}
\end{algorithm}

For each time step $l$, each SSM selective block generates input-dependent vectors \( \bm{b}_l \) and \( \bm{c}_l \) of dimension \( N>D\), along with a step size vector $\bm{\Delta}_l$ of dimension $D$. Each entry $\Delta_{l,d}$ of vector $\bm{\Delta}_l$ correspond to the step size used by the $d$-th parallel model. Following \cite{gu2023mamba}, vector \( \bm{b}_l \) and \( \bm{c}_l \) are obtained via a trainable linear function $\text{Linear}_N(\cdot)$  applied to the input token vector $\bm{u}_l$. Furthermore, vector $\bm{\Delta}_l$ is obtained as the output of the $\text{Softplus}(\cdot)$ operation applied to the sum of a bias vector $\epsilon$ and of the output of a trainable scalar linear function $\text{Linear}_1(\cdot)$ applied to input $\bm{u}_l$. 

The $l$-th output token \( \bm{o}_l \) of a former layer serves as the $l$-th input token \( \bm{u}_l \) of the next layer. The output sequence $\bm{o}_1,\ldots,\bm{o}_L$ from the last layer serves as the final output of the network.

\subsection{Pre-Training}
The output of the $N_{\text{layer}}$-th SSM block at the $L$-th step $\bm{o}_L$ is mapped to a complex vector \( \hat{x}_{k|\theta} \), corresponding to the estimate of the transmitted data symbol \( x_k \), where $\theta$ represents the collection of the trainable parameter of the SSM-ICL equalizer. The SSM parameter vector $\theta$  is optimized offline using a pre-training task set \( \mathcal{T}_{\text{tr}} \) comprising \( N_{\text{tr}} \) tasks sampled i.i.d. from an unknown task distribution dictated by the specific deployment and network conditions. 

For each task $\tau\in\mathcal{T}_{\text{tr}}$, we assume access to \( N_{\text{ex}} \) independently generated contexts \( \{\mathcal{C}_{\tau,k} \}^K_{k=1}\) and a test pairs \( \{(x_k, \tilde{\bm{y}})\}^K_{k=1} \). Using these data, during the pre-training phase, the parameter $\theta$ is optimized to minimize the mean squared error averaged over the pre-training tasks \( \mathcal{T}_{\text{tr}} \), i.e.,
\begin{equation}
    \mathcal{L}(\theta)= \sum_{\tau \in \mathcal{T}_{\mathrm{tr}}} \sum_{k=1}^{K} \mathbb{E} \left[ \left\| \hat{x}_{k|\theta} - x_k \right\|^2 \right],
\end{equation}
where the expectation is calculated over the \( N_{\mathrm{ex}} \) examples available for each task \( \tau \).

\section{Experiments}
In this section, we evaluate the performance of the SSM-ICL equalizer for channel equalization in a cell-free multi-user MIMO system with limited fronthaul capacity. We compare its performance to the transformer-based ICL (T-ICL) equalizer proposed in \cite{zecchin2024cell} and the to conventional centralized linear minimum mean squared error (LMMSE) estimator \cite{bjornson2019making}.
\subsection{Cell-Free MIMO Setup}
We consider a cell-free MIMO architecture consisting of $M=4$ APs, in which each AP is equipped with $N_{\text{ant}}=2$ antennas. $K=4$ UEs are distributed randomly in a $1 \text{ km}\times 1 \text{ km}$ area. Between 2 and 4 randomly selected UEs reuse the same pilot sequence, leading to pilot contamination.  Communication takes place at a carrier frequency of 2 GHz, with path loss and fading characteristics modeled according to the 3GPP Urban Microcell standard.

As in \cite{bjornson2019making}, pilot sequences of length $T_{\text{P}}= 8$ are generated using a Walsh-Hadamard matrix. During data transmission, The $k$-th UE picks the modulation schemes at random from the set \{BPSK, 8-PSK, 4-QAM, 16-QAM, 64-QAM\}. To account for the limited capacity of the fronthaul, each AP applies $b$-bit scalar quantization to the received signal, which is designed to minimize the average mean squared quantization error as in \cite{zecchin2024cell}.

\subsection{Mean Squared Error Performance}

In this subsection, we examine the performance in terms of the MSE of the equalized symbols.
We consider three baselines: (\emph{i}) the centralized LMMSE equalizer studied in \cite{bjornson2019making} assuming unlimited fronthaul capacity ($b = \infty$), (\emph{ii}) the centralized LMMSE equalizer operating under fronthaul capacity constraints through the mentioned fronthaul scaler quantizer $\mathcal{Q}_b(\cdot)$, and (\emph{iii}) The 6-layer T-ICL equalizer described in \cite{zecchin2024cell}, which features an embedding dimension of 64 and 8 attention heads per layer. For the proposed SSM-ICL equalizer, we employ a 6-layer Mamba network with a hidden dimension of 64. We train the SSM-ICL and T-ICL models separately for each $b\in\{1,\ldots,8\}$ at $\text{SNR}=24$ dB on a training dataset with \( N_{\text{tr}} =8192\) tasks.

\begin{figure}[t]
    \centering
    \includegraphics[width=1\linewidth]{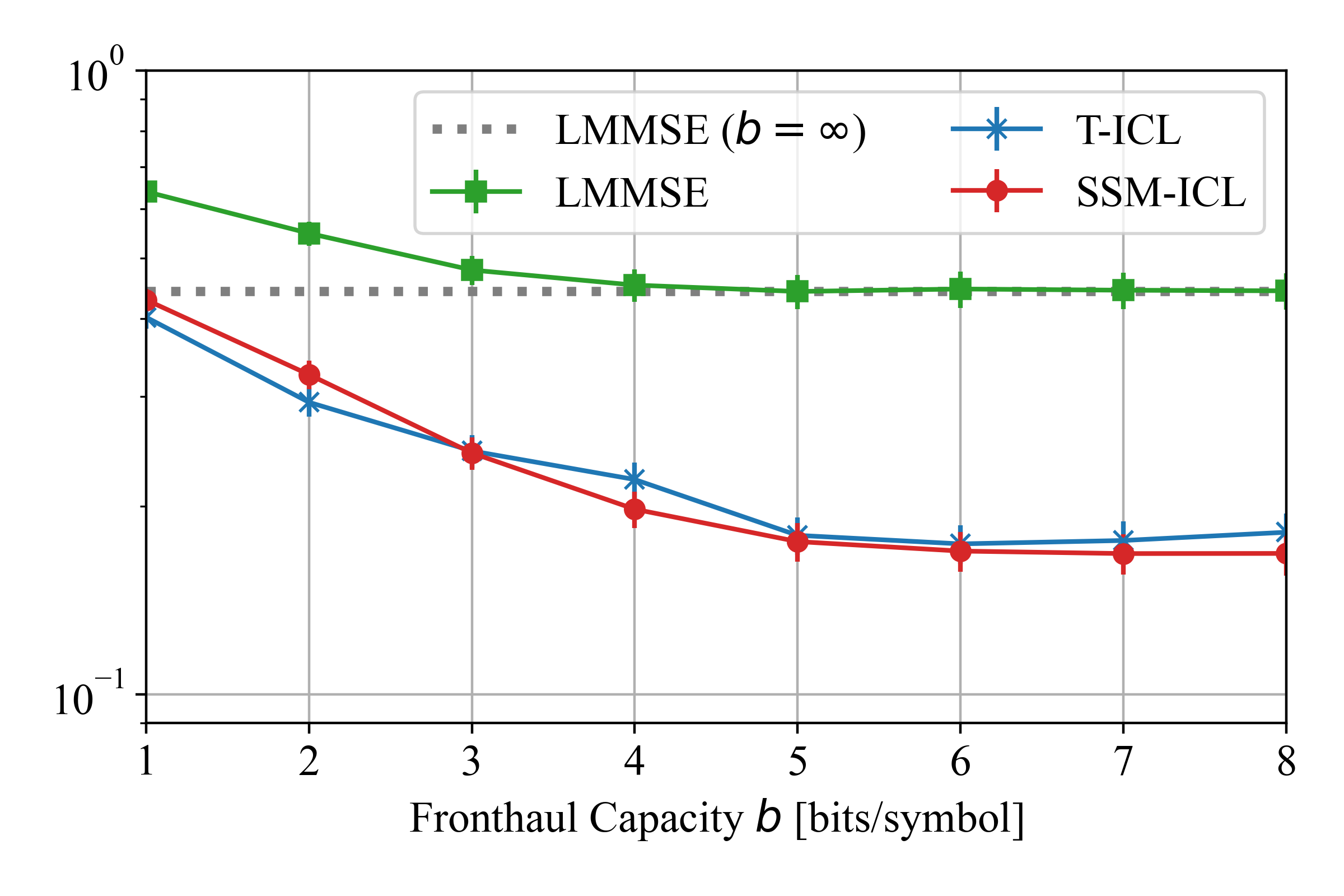}
    \caption{MSE versus the fronthual capacity (SNR=24 dB)}
    \label{fig:mse_bits}
\end{figure}

\begin{figure}[t]
    \centering
    \includegraphics[width=1\linewidth]{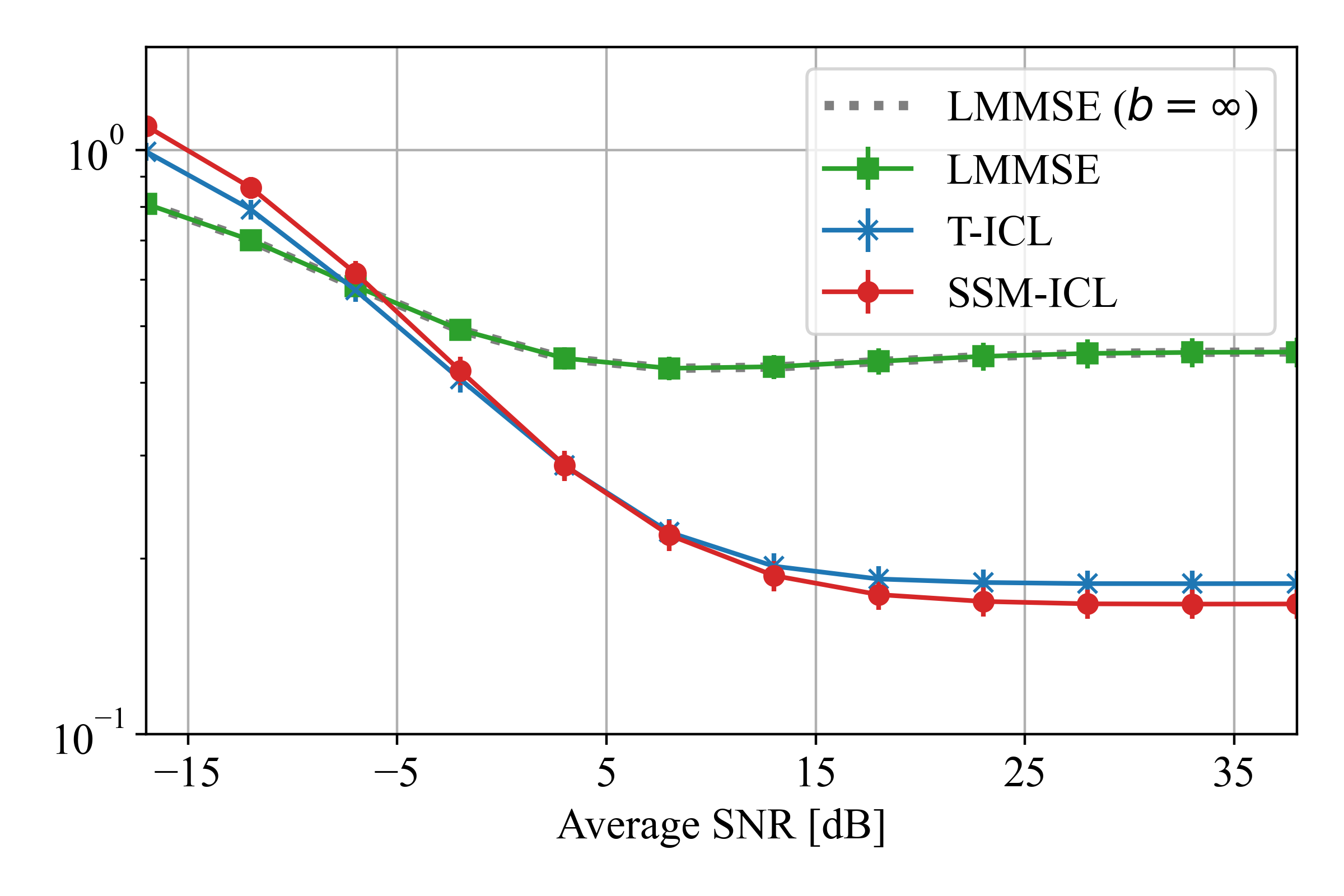}
    \caption{MSE versus signal SNR (fronthual capacity $b=8$).}
    \label{fig:mse_snr}
\end{figure}

 Fig. \ref{fig:mse_bits} presents the MSE for an SNR of 24 dB as a function of the fronthaul capacity $b$. Due to the limited fronthaul capacity, for centralized LMMSE equalizer exhibits poor performance when the quantization level is below $b = 4$ bits, while, as $b$ increases, the MSE gradually approaches that of the LMMSE benchmark with unlimited fronthaul capacity. Both the T-ICL and SSM-ICL equalizers significantly outperform the LMMSE solution, even surpassing the infinite-capacity LMMSE benchmark for $b \geq 3$ bits. This performance gain is attributed to the capacity of ICL equalizers to adapt to different fronthaul capacity levels. The SSM-ICL equalizer performs comparably to the T-ICL, indicating that SSMs can effectively capture contextual information without the need for processing all previous tokens simultaneously.

In Fig. \ref{fig:mse_snr}, we display the MSE as a function of the SNR for a fixed fronthaul capacity of $b = 8$ bits. The centralized LMMSE equalizer performs poorly, with an MSE exceeding 0.4 due to pilot contamination. In contrast, both ICL equalizers achieve MSE values below 0.2 at high SNRs, with the Mamba model slightly outperforming the transformer-based counterpart. This demonstrates the superior capability of the SSM-based ICL model to use contextual information to manage pilot contamination.

\subsection{Computational Efficiency}
\begin{figure}
    \centering
    \includegraphics[width=1\linewidth]{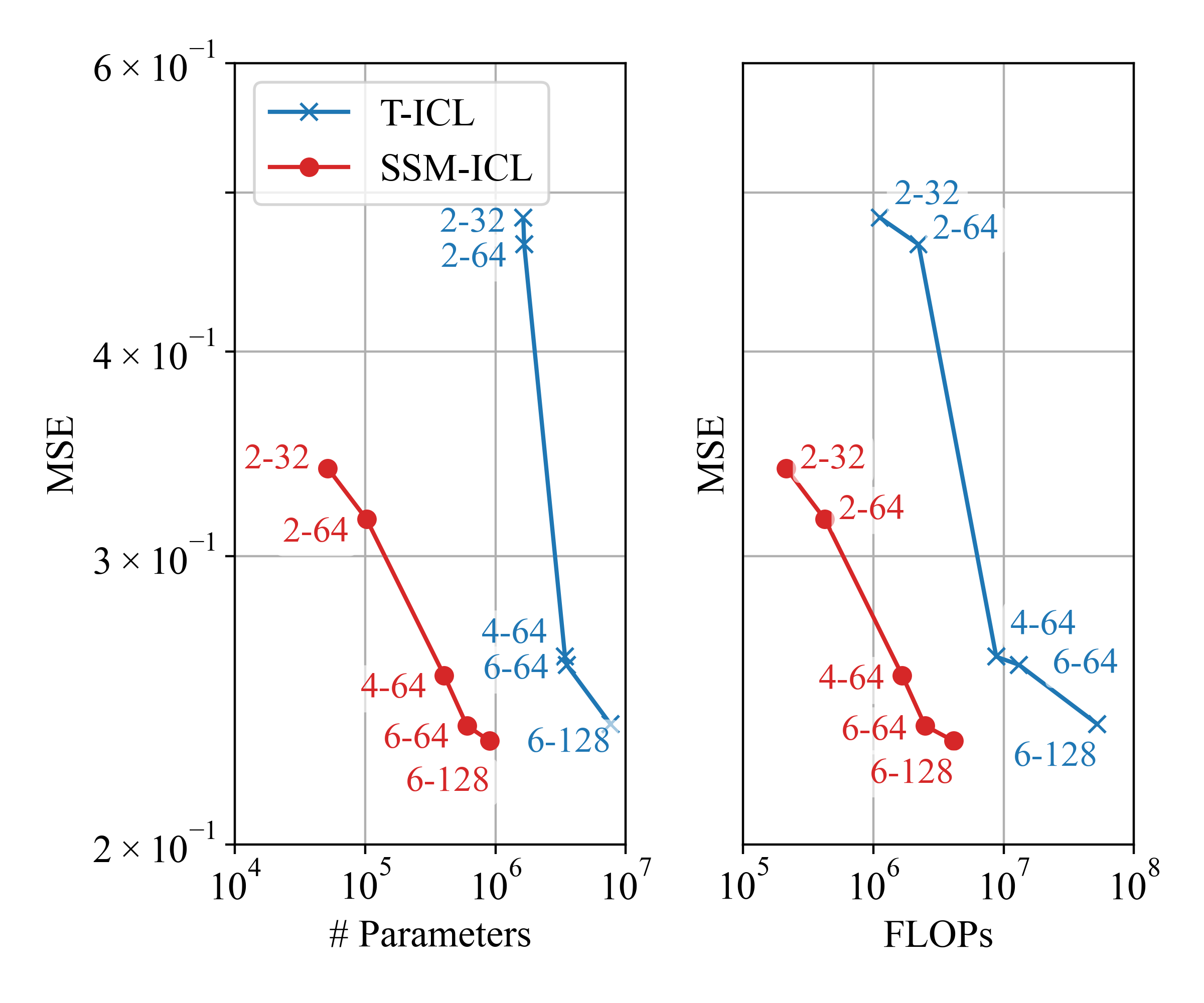}
    \caption{In the left panel, the MSE of the T-ICL and SSM-ICL equalizers as a function of the number of model parameters. In the right panel, the MSE as a function of the number of FLOPs required to decode a data symbol.  Data point labels represent the number of layers and hidden-state dimensions, respectively.}
    \label{fig:model_size}
\end{figure}

The key advantage of SSMs lies in their computational efficiency, which is achieved by processing contextual information sequentially rather than in parallel, as transformers do. To quantify the potential computational benefits of SSMs for ICL-based equalization, in this subsection, we study the MSE as a function of the model size and of the number of floating-point operations (FLOPs) necessary to decode a data symbol. This is done by scaling the number parameters of the SSM-ICL and T-ICL models while keeping the same number of layers $ N_\text{layer}$ and hidden state dimension $N$. The fronthaul capacity is set to $b=8$ bits and the $\text{SNR}=24$ dB.

As shown in Fig. \ref{fig:model_size}, as the number of parameters increases, both the T-ICL and the SSM-ICL achieve smaller MSE, with SSM-ICL demonstrating a consistently higher performance across all tested scenarios. For instance, for a target MSE of approximately 0.23, SSM-ICL ($N_\text{layer}=4$, $N=64$) requires $8.5\times$ fewer parameters and $5.5\times$ fewer FLOPs than T-ICL with the same layer count and hidden state dimensions.

Another notable finding is that SSM-ICL achieves significantly lower MSE with improved performance when instantiated with extremely few parameters, as in the case of configurations with 2 layers. For example, the T-ICL equalizer ($N_\text{layer}=2$, $N=32$) achieves an MSE equal to 0.48, while the SSM-ICL equalizer with the same configuration attains an MSE equal to 0.33. This performance advantage of SSM-ICL at reduced scales suggests that SSM-ICL better preserves essential information in the context even when operating under a constrained parameter budget. This behavior underlines the scalability of SSM-ICL, making it a more effective choice in resource-limited settings, where both performance and model compactness are critical.

\section{Conclusion}
This work explores introduces state-space models (SSMs) for in-context learning (ICL)-based channel equalization, focusing on cell-free massive MIMO systems with constrained fronthaul capacity. Unlike transformer-based ICL equalizers, which rely on batch processing and incur high computational and memory costs, SSMs offer a resource-efficient alternative, substantially reducing both memory and computational demands. Numerical results show that SSM-based ICL equalizers outperform traditional estimation methods in cell-free MIMO environments, particularly under limited fronthaul capacity and pilot contamination. Additionally, the proposed SSM-based ICL equalizers achieve comparable mean-square error performance to transformer-based ICL equalizers, while using approximately eight times fewer parameters and five times fewer operations per symbol. These findings highlight the potential of SSMs as a practical, efficient solution for real-time equalization in next-generation open radio access network architectures.

\bibliography{reference}

\begin{thebibliography}{10}
\providecommand{\url}[1]{#1}
\csname url@samestyle\endcsname
\providecommand{\newblock}{\relax}
\providecommand{\bibinfo}[2]{#2}
\providecommand{\BIBentrySTDinterwordspacing}{\spaceskip=0pt\relax}
\providecommand{\BIBentryALTinterwordstretchfactor}{4}
\providecommand{\BIBentryALTinterwordspacing}{\spaceskip=\fontdimen2\font plus
\BIBentryALTinterwordstretchfactor\fontdimen3\font minus \fontdimen4\font\relax}
\providecommand{\BIBforeignlanguage}[2]{{%
\expandafter\ifx\csname l@#1\endcsname\relax
\typeout{** WARNING: IEEEtran.bst: No hyphenation pattern has been}%
\typeout{** loaded for the language `#1'. Using the pattern for}%
\typeout{** the default language instead.}%
\else
\language=\csname l@#1\endcsname
\fi
#2}}
\providecommand{\BIBdecl}{\relax}
\BIBdecl

\bibitem{bonati2021intelligence}
L.~Bonati, S.~D'Oro, M.~Polese, S.~Basagni, and T.~Melodia, ``{Intelligence and learning in O-RAN for data-driven NextG cellular networks},'' \emph{IEEE Communications Magazine}, vol.~59, no.~10, pp. 21--27, 2021.

\bibitem{balasubramanian2021ric}
B.~Balasubramanian, E.~S. Daniels, M.~Hiltunen, R.~Jana, K.~Joshi, R.~Sivaraj, T.~X. Tran, and C.~Wang, ``{RIC: A RAN intelligent controller platform for AI-enabled cellular networks},'' \emph{IEEE Internet Computing}, vol.~25, no.~2, pp. 7--17, 2021.

\bibitem{hou2024automaticaimodelselection}
\BIBentryALTinterwordspacing
Q.~Hou, M.~Zecchin, S.~Park, Y.~Cai, G.~Yu, K.~Chowdhury, and O.~Simeone, ``Automatic {AI} model selection for wireless systems: Online learning via digital twinning,'' 2024. [Online]. Available: \url{https://arxiv.org/abs/2406.15819}
\BIBentrySTDinterwordspacing

\bibitem{garg2022can}
S.~Garg, D.~Tsipras, P.~S. Liang, and G.~Valiant, ``What can transformers learn in-context? a case study of simple function classes,'' \emph{Advances in Neural Information Processing Systems}, vol.~35, pp. 30\,583--30\,598, 2022.

\bibitem{zecchin2023context}
M.~Zecchin, K.~Yu, and O.~Simeone, ``In-context learning for {MIMO} equalization using transformer-based sequence models,'' in \emph{2024 IEEE International Conference on Communications Workshops (ICC Workshops)}, 2024, pp. 1573--1578.

\bibitem{rajagopalan2023transformers}
V.~Rajagopalan, V.~T. Kunde, C.~S.~K. Valmeekam, K.~Narayanan, S.~Shakkottai, D.~Kalathil, and J.-F. Chamberland, ``Transformers are efficient in-context estimators for wireless communication,'' \emph{arXiv preprint arXiv:2311.00226}, 2023.

\bibitem{zecchin2024cell}
M.~Zecchin, K.~Yu, and O.~Simeone, ``Cell-free multi-user {MIMO} equalization via in-context learning,'' in \emph{2024 IEEE 25th International Workshop on Signal Processing Advances in Wireless Communications (SPAWC)}, 2024, pp. 646--650.

\bibitem{song2024neuromorphic}
Z.~Song, O.~Simeone, and B.~Rajendran, ``Neuromorphic in-context learning for energy-efficient {MIMO} symbol detection,'' in \emph{2024 IEEE 25th International Workshop on Signal Processing Advances in Wireless Communications (SPAWC)}, 2024, pp. 1--5.

\bibitem{abbas2024leveraging}
M.~Abbas, K.~Kar, and T.~Chen, ``Leveraging large language models for wireless symbol detection via in-context learning,'' \emph{arXiv preprint arXiv:2409.00124}, 2024.

\bibitem{song2024transformer}
Z.~Song, Y.~Ma, C.~You, H.~Yuan, J.~Peng, and Y.~Gao, ``Transformer-based adaptive {OFDM MIMO} equalization in intelligence-native {RAN},'' in \emph{2024 IEEE/CIC International Conference on Communications in China (ICCC Workshops)}.\hskip 1em plus 0.5em minus 0.4em\relax IEEE, 2024, pp. 179--184.

\bibitem{chen2023learning}
L.~Chen, S.~T. Jose, I.~Nikoloska, S.~Park, T.~Chen, O.~Simeone \emph{et~al.}, ``Learning with limited samples: {M}eta-learning and applications to communication systems,'' \emph{Foundations and Trends{\textregistered} in Signal Processing}, vol.~17, no.~2, pp. 79--208, 2023.

\bibitem{simeone2022machine}
O.~Simeone, \emph{{M}achine {L}earning for {E}ngineers}.\hskip 1em plus 0.5em minus 0.4em\relax {C}ambridge {U}niversity{ P}ress, 2022.

\bibitem{nikoloska2022modular}
I.~Nikoloska and O.~Simeone, ``Modular meta-learning for power control via random edge graph neural networks,'' \emph{IEEE Transactions on Wireless Communications}, vol.~22, no.~1, pp. 457--470, 2022.

\bibitem{quek2017cloud}
T.~Quek, M.~Peng, and O.~Simeone, \emph{Cloud {R}adio {A}ccess {N}etworks: {P}rinciples, {T}echnologies, and {A}pplications}.\hskip 1em plus 0.5em minus 0.4em\relax Cambridge {U}niversity {P}ress, 2017.

\bibitem{nayebi2015cell}
E.~Nayebi, A.~Ashikhmin, T.~L. Marzetta, and H.~Yang, ``Cell-free massive {MIMO} systems,'' in \emph{2015 49th Asilomar Conference on Signals, Systems and Computers}, 2015, pp. 695--699.

\bibitem{bjornson2019making}
E.~Bj{\"o}rnson and L.~Sanguinetti, ``Making cell-free massive {MIMO} competitive with {MMSE} processing and centralized implementation,'' \emph{IEEE Transactions on Wireless Communications}, vol.~19, no.~1, pp. 77--90, 2019.

\bibitem{ibrahim2021uplink}
M.~Ibrahim, S.~Elhoushy, and W.~Hamouda, ``Uplink performance of mm{W}ave-fronthaul cell-free massive {MIMO} systems,'' \emph{IEEE Transactions on Vehicular Technology}, vol.~71, no.~2, pp. 1536--1548, 2021.

\bibitem{masoumi2019performance}
H.~Masoumi and M.~J. Emadi, ``Performance analysis of cell-free massive {MIMO} system with limited fronthaul capacity and hardware impairments,'' \emph{IEEE Transactions on Wireless Communications}, vol.~19, no.~2, pp. 1038--1053, 2019.

\bibitem{Vaswani2017AttentionIA}
A.~Vaswani, N.~Shazeer, N.~Parmar, J.~Uszkoreit, L.~Jones, A.~N. Gomez, {\L}.~Kaiser, and I.~Polosukhin, ``Attention is all you need,'' in \emph{Advances in neural information processing systems}, vol.~30, 2017.

\bibitem{gu2021efficiently}
A.~Gu, K.~Goel, and C.~R{\'e}, ``Efficiently modeling long sequences with structured state spaces,'' \emph{arXiv preprint arXiv:2111.00396}, 2021.

\bibitem{gu2023mamba}
A.~Gu and T.~Dao, ``Mamba: Linear-time sequence modeling with selective state spaces,'' \emph{arXiv preprint arXiv:2312.00752}, 2023.

\bibitem{sushma2024state}
N.~M. Sushma, Y.~Tian, H.~Mestha, N.~Colombo, D.~Kappel, and A.~Subramoney, ``State-space models can learn in-context by gradient descent,'' \emph{arXiv preprint arXiv:2410.11687}, 2024.

\bibitem{park2024can}
J.~Park, J.~Park, Z.~Xiong, N.~Lee, J.~Cho, S.~Oymak, K.~Lee, and D.~Papailiopoulos, ``Can mamba learn how to learn? a comparative study on in-context learning tasks,'' \emph{arXiv preprint arXiv:2402.04248}, 2024.

\bibitem{demir2024cell}
{\"O}.~T. Demir, M.~Masoudi, E.~Bj{\"o}rnson, and C.~Cavdar, ``Cell-free massive {MIMO} in {O-RAN}: Energy-aware joint orchestration of cloud, fronthaul, and radio resources,'' \emph{IEEE Journal on Selected Areas in Communications}, 2024.

\bibitem{3gpp_split}
\BIBentryALTinterwordspacing
``{Study on new radio access technology: Radio access architecture and interfaces, version 14.0.0,},'' \emph{3rd Gener. Partnership Project (3GPP), Sophia Antipolis, France, Rep. (TR) 38.801,}, Apr. 2017. [Online]. Available: \url{http://www.3gpp.org/DynaReport/38801.htm}
\BIBentrySTDinterwordspacing

\end{thebibliography}

\end{document}